\documentclass[pss]{wiley2sp} 
\usepackage{amsmath}
\begin{document}

% Title of the article
\title{From acene to graphene spectrum of $\pi$ electrons with the use of the Green's function}

% Abbreviated title for the page headers
\titlerunning{From acene to graphene spectrum \ldots}

% Authors
\author{%
  Lyuba Malysheva\textsuperscript{\textsf{%\bfseries 1
}} and
  Alexander Onipko\textsuperscript{\textsf{%\bfseries 2,
\Ast}} }
  %Third Author\textsuperscript{\textsf{\bfseries 3}}, \ldots}

% Abbreviated list of authors for the page haeders
\authorrunning{Lyuba Malysheva et al.}

%E-mail-address of corresponding author
\mail{e-mail
  \textsf{aleon@ifm.liu.se}, Phone
  +46-13-288904, Fax +46-13-28 8969}

% author's affiliations/addresses
\institute{%
%  \textsuperscript{1}\,
Bogolyubov Institute for Theoretical Physics, 03680, Kiev, Ukraine}
%\\
%  \textsuperscript{2}\,Division of Molecular Physics,
%Department of Physics,
%Link{\"o}ping University,
%S-581 83 Link{\"o}ping, Sweden}
% \\  \textsuperscript{3}\,Third address}

\received{XXXX, revised XXXX, accepted XXXX} % do not change, will be filled in by the publisher
\published{XXXX} % do not change, will be filled in by the publisher

%Please select four to six PACS-codes from the enclosed list (PACS.txt) or from www.aip.org/pacs)
 % For example: 71.20.Ps

\abstract{Origin of the spectrum of $\pi$ electrons that results from coupling of $N$ $\cal N$-long acenes via C-C covalent bonding have been traced with the use of the Green function formalism. Exact expressions of acene and graphene Green's functions, which are useful for analysis of electronic properties of these macromolecules, are obtained and advanced to the form suitable for instructive applications.}

\maketitle   % please do not remove

\section{Introduction.}
Free standing graphene, a monoatomic layer of graphite is a new material \cite{Nov1,Nov2} that exhibits exceptional electronic quality 
\cite{Nov1,Nov2,Nov3,Zhang,Geim} and a unique nature of charge carriers. On one hand, electrons and holes appear as quasi-particles in condensed-matter physics when the Schr\"odinger equation directs our prediction of their behavior \cite{Wallace,Fujita,Nakada,Saito,W1,W2,Thomsen,Peres1,prb2007,W3,Lyuba1}. On the other hand, they behave as relativistic particles which are subject to the Dirac equation \cite{Sem,Khv,Ando,Brey,Gus1,Gus2}. Much theoretical effort has been focused primarily on the electronic states of infinite graphene. In the continuum approximation, these states can be mapped into the Hamiltonian of 2+1 dimensional quantum electrodynamics with Dirac fermions. This description is based on the cone-like spectrum near the neutrality (zero-energy) points of 2D graphite band structure, see \cite{Gus2} and references therein. Less attention has been paid to graphene as a molecular species. From this point of view, graphene can be considered as a macromolecule obtained by C-C bonding of acene chains as shown in Fig. 1. In certain aspects, the macromolecule model provides a better approximation for the description of real graphene structures \cite{Lyuba1}, particularly, in the nanometer scale, where the continuous approximation is of a limited value \cite{Lyuba2}. 

\begin{figure}[htb]%
\centering
\includegraphics*[width=0.8\linewidth,height=0.63\linewidth]{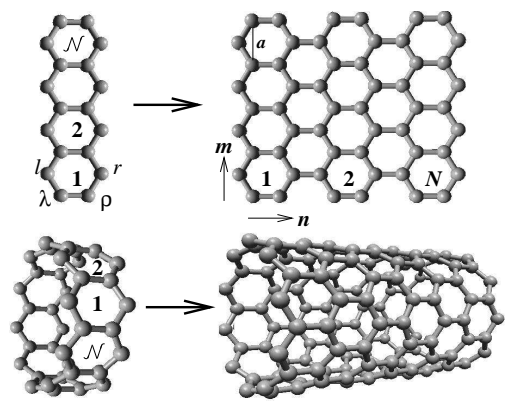}
\caption{$\cal N$-long linear acene as a basic unit of $N$$\times$$\cal N$ sheet of graphene honeycomb lattice (up) and $\cal N$-long cyclacene playing the same role for zigzag ($\cal N$,0) carbon nanotubes (CNTs).}
\label{Fig1}
\end{figure}

As is clear from Fig.~1, $\cal N$-long linear acenes are basic units of $N$$\times$$\cal N$ sheet of graphene honeycomb lattice, whereas $\cal N$-long cyclacenes play the same role for zigzag ($\cal N$,0) carbon nanotubes (CNTs). In this sense, the graphene and acene spectra are intimately interrelated. The electronic structure of acenes, the semi-empirical description of which goes back to the Pariser and Parr classic works \cite{PP,P}, is understood fairly well \cite{Hoff,Kiv,Houk}. This study elucidates the relationship between the graphene and acene spectra. Keeping in mind subsequent applications, our analysis has been performed with the use of the Green's function method. All derivations are based on the standard operator equation 
\begin{equation}\label{1}
G^{\rm M} =\left[EI-H^{M}\right]^{-1},
\end{equation}
where $H^{\rm M}$ is the tight-binding Hamiltonian operator of acene (M=A) or graphene (M=G); the electron energy $E$ is in unites of $|t|$, $t$ denotes C-C hopping integral. For the Green's function matrix elements, notations $\langle m\alpha|G^{\rm A}|m'\beta\rangle$ $\equiv$ $G^{\rm A}_{m\alpha,m'\beta}$ and $\langle mn\alpha |G^{\rm G}|m'n'\beta\rangle$ $\equiv$ $G^{\rm G}_{mn\alpha,m'n'\beta}$ will be used for acene and graphene, respectively;  $|m\alpha\rangle$ and $|mn\alpha\rangle$ denote $2p_z$ orbitals at the $\alpha$th carbon atom of the $m$th and $mn$th hexagon, respectively, as shown in Fig.~1. The formal description of acenes and cyclacenes, and gra\-phe\-ne and CNT is very similar. However, in the interests of simplicity, only acene--graphene pair will be in focus. An extended discussion that also includes cyclacenes and zigzag CNTs can be found in Ref.~\cite{Lyuba4}

\section{Acene Green's function.}
For linear acenes, the matrix form of solution to Eq.~(\ref{1}) is given by \cite{Klym1}
\begin{equation} \label{6} 
 G^{\rm A}_{m \alpha,m'\beta} = \frac{2}{{\cal N}+1}\sum_{j=1}^{\cal N}
g_{\alpha,\beta}^j \sin(\xi_j m)\sin(\xi_j m'),
\end{equation}
where $\xi_j= \pi j/({\cal N}{\rm +}1)$, $j=1,2,...,\cal N$, and
\begin{equation} \label{7} 
{\cal D}_jg_{\alpha,\beta}^j =
\left\{
\begin{array}{lr}
E\left[E^2{\rm -}4\cos^2(\xi_j/2){\rm -}1\right ],&
\alpha =\beta=l,r,\\
4\cos^2(\xi_j/2),&\hspace{-0.9cm}
\alpha =l(r), \beta=r(l),\\
\end{array}
\right.
\end{equation}
where ${\cal D}_j=[E^2-4\cos^2(\xi_j/2)]^2-E^2$. For the rest of matrix elements, we obtain
$$
G^{\rm A}_{m \lambda,m'
\left\{ \begin{smallmatrix} \lambda \\ \rho \end{smallmatrix} \right \}}
 =\frac{\delta_{m,m'} \left\{ \begin{smallmatrix} E \\ 1 \end{smallmatrix} \right \}}{E^2-1}+ \frac{8}{{\cal N}+1}
$$
\begin{equation} \label{8}
\times\sum_{j=1}^{\cal N}
g_{\alpha,\beta}^j \cos^2(\xi_j/2)\sin[\xi_j (m{\rm -}1/2)]\sin[\xi_j (m'{\rm -}1/2)],
\end{equation}
$$
{\cal D}_j (E^2-1)g_{\alpha,\beta}^j
$$
\begin{equation} \label{9} 
 =
\left\{
\begin{array}{lr}
E[E^2{\rm -}4\cos^2(\xi_j/2){\rm +}1],&
\alpha =\beta=\lambda,\rho,\\
2E^2{\rm -}4\cos^2(\xi_j/2),&  
\alpha =\lambda(\rho),  \beta=\rho(\lambda),\\
\end{array}
\right.
\end{equation}
and
$$
 G^{\rm A}_{m \alpha,m'\beta}
$$
\begin{equation} \label{10} 
 = \frac{4}{{\cal N}+1}
\sum_{j=1}^{\cal N}g_{\alpha,\beta}^j\cos(\xi_j/2) \sin[\xi_j (m{\rm -}1/2)]\sin(\xi_j m'),
\end{equation}
\begin{equation} \label{11} 
{\cal D}_j g_{\alpha,\beta}^j =
\left\{
\begin{array}{lr}
E^2-4\cos^2(\xi_j/2),&
\alpha =\lambda(\rho),  \beta=l(r),\\
E,&
\alpha =\lambda(\rho), \beta=r(l).\\
\end{array}
\right.
\end{equation}

As shown in \cite{Lyuba1}, defined in Eq. (\ref{7}) quantities $g^j_{l,l}=g^j_{r,r}$ and $g^j_{l,r}=g^j_{r,l}$ determine the dispersion relation for an $N$$\times$${\cal N}$  sheet of graphene $E(\kappa_j,\xi_j)$ via an equation
\begin{equation} \label{12} 
2 g_{l,r}^j \cos \kappa_j = 1-(g_{l,l}^j)^2 +(g_{l,r}^j)^2,
\end{equation}
where $\kappa_j$ is the second, complementing $\xi_j$ quantum number. $2N$ values of $\kappa^\pm_{j,\nu}$, $\nu$ = $0,1,...,N-1$, for each $j$ classify the graphene spectrum into 4${\cal N}N$ $j,\nu$ levels; see below.  

\begin{figure}[htb]%
\centering
\includegraphics*[width=0.9\linewidth,height=0.6\linewidth]{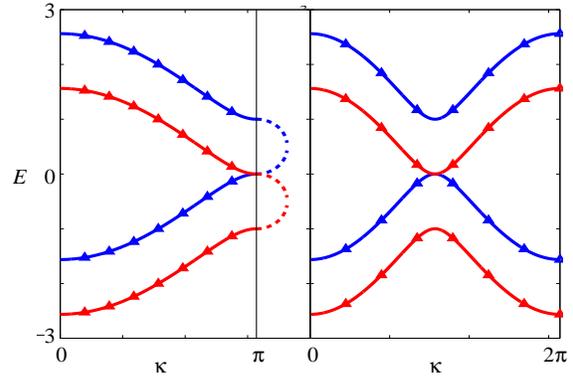}
\caption{Band structure of polyacene, $E(\kappa{\rm =}k^+)$ (red) and $E(\kappa{\rm =}k^-)$ (blue), calculated from Eq.~(\ref{13}) for infinite linear acene (left panel) and cyclacene (right panel). Continuation to imaginary values of wave vectors, $k^\pm\rightarrow \pm\pi\pm i\delta$, $0\leq\delta\leq 2 \ln((1+\sqrt{17})/4)\approx 0.5$, is shown by dashed lines. Solid lines of the right panel for $\kappa\le\pi$ repeats similar calculations of Ref.~\cite{Kiv}. Triangles show energy levels for $\cal N$=7 according to Eq.~(\ref{15}).}
\label{Fig2}
\end{figure}

By performing summation over $j$, it is possible to express the energy dependence of acene Green's function in terms of (dimensionless) wave vectors $k^+$ and $k^-$, subjected to equation
\begin{equation}\label{13}
4\cos^2(k^\pm/2)=E(E \pm 1).
\end{equation}
For example, $G^{\rm A}_{1\lambda(\rho),({\cal N}+1)\lambda(\rho)}= G_{k^+} +G_{k^-} $, where
\begin{equation}\label{14}
G_{k^\pm} =
\frac{1}{2(E\pm1)}\frac{\sin k^\pm}{ \sin({\cal N}+1)k^\pm}.
\end{equation}
Other matrix elements have similar expressions. As seen from Fig. 2, for the given energy, either both $k^+$ and $k^-$ are real, or one is real, whereas the other one is imaginary. In the latter case, matrix elements which refer to the opposite sites of the chain have two typical terms: One is exponential, $\sim \exp\left(-\sqrt{|E|}{\cal N}\right )$, and the second term has a singular character. Hence, the proba\-bi\-lity of electron transmission through acenes ($\sim G^{\rm A\,2}_{1,({\cal N}{\rm +}1)}$ \cite{Chapt}) can occur via partly coherent and partly tunneling mechanisms. Tunneling governed by a single exponential factor, as it takes place in conjugated oligomers \cite{Chapt,Mag}, is not possible. This is a reflection of a metallic nature of polyacenes \cite{Hoff,Kiv}.

\begin{figure*}[htb]%
\includegraphics*[width=0.96\linewidth,height=0.24\linewidth]{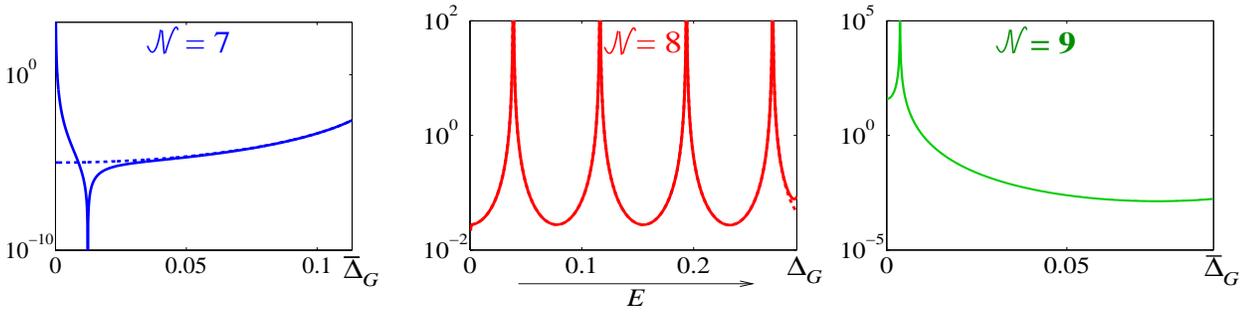}
\caption{ 
Green's function matrix element squared, $G^{\rm G\,2}_{11l,1Nr}$, $N$=20, for semiconducting (blue, green) and metallic (red) graphene ribbons; the correspondance between colors and $\cal N$ values is the same, as in Fig.~4. An increase/decrease of $N$ results in a shift of the deep and peak in the left and right panels, respectively, towards smaller/larger energies and also, in a larger/smaller number of peaks in the mid panel. Differences between the exact Green's function expansion and its approximation by the $j^*$th term (mid and right panels), and by a sum of the $j^*$th and ($j^*$+1)th terms of the expansion (left panel) are vanishingly small. Dashed blue line shows an approximation by the $j^*$ term. Notations $\bar\Delta_{_{\rm G}} =\Delta_{_{\rm G}}/3 =E_g/2=\pi/[2\sqrt{3}({\cal N}{\rm +1})]$ are explained in the text.}
\label{Fig3}
\end{figure*}

\subsection{Acene electron spectrum.}
The acene electronic structure is determined by equation $(E^2-1){\cal D}_j=0$; zeros of ${\cal D}_j$ give a four-band spectrum 
\begin{equation} \label{15} 
E_{c(v)}^{\pm} =+(-) \frac{1}{2}\left[\mp1\pm \sqrt{1+16\cos^2(\xi_j/2)}\right],
\end{equation}
where each of two conduction ($c$) and two valence ($v$) bands has $\cal N$ levels. Note that two extra levels $E=\pm1$ do not appear in Eq.~(\ref{6}) as poles of the acene Green's function. The states with these energies correspond to zero wave function amplitudes at $l$ and $r$ sites and thus, break the acene chain into $\lambda$-$\rho$ "isolated" pairs.

For long chains ${\cal N}$$>>$1 and $|E|<<1$, spectrum (\ref{15}) is well approximated by $
E=\pm q_\mu^2$ with $q_\mu$$=\pi\mu/{\cal N}$, $\mu$ = 1,2,... . 
In the limit ${\cal N}$$\rightarrow$$\infty$, the quantum number $\xi_j$ can be considered as a continuous variable, $\xi_j\rightarrow k$, and $q_\mu\rightarrow q =\pi{\rm -} k$ acquires the meaning of $k$ separation from the zero-energy point at $k=\pi$. The spectrum of polyacene (an infinite acene chain) that comes out from the spectra of a linear acene and cyclacene, is shown in Fig.~2. 

\section{Graphene Green's function.}

Similarly to Eq.~(\ref{6}), matrix elements of graphene Green's function can be represented in the form of an expansion 
\begin{equation} \label{16} 
G^{\rm G}_{mn\alpha,m'n'\beta} = 
\frac{2}{{\cal N}+1}\sum_{j=1}^{\cal N} G^j_{n\alpha,n'\beta} 
\sin(\xi_j m)\sin(\xi_j m'), 
\end{equation}
where $G^j_{n\alpha,n'\beta} = G^{j+}_{n\alpha,n'\beta} +G^{j-}_{n\alpha,n'\beta}$. 
Here, we restrict ourselves by representative examples of matrix elements refering to the $l$ and $r$ sites, namely,
\begin{equation} \label{17} 
 G^{j\pm}_{1l,
\left\{ \begin{smallmatrix} 1 \\ N \end{smallmatrix} \right \}
\left\{ \begin{smallmatrix} l \\ r \end{smallmatrix} \right \}}  
= 
\frac{
\left\{ \begin{smallmatrix} \displaystyle g_{l,l}^j \sin (\kappa^\pm_{j} N)\\
 \displaystyle  g_{l,r}^j \sin \kappa^\pm_{j}  \end{smallmatrix} \right \}} 
{\sin(\kappa^\pm_{j} N)-g_{l,r}^j\sin[\kappa^\pm_{j} (N-1)]},
\end{equation}
where labels + and $-$ correspond to, respectively, "plus" and "minus" branches of dispersion relation (\ref{12}). This relation can be rewritten as 
$$E_{c(v)}^{j\pm} =$$ 
\begin{equation}\label{18}
+(-) \sqrt{1+4\cos^2 (\xi_j/2) \pm 4|\cos (\xi_j/2)  \cos (\kappa_j/2)|}.
\end{equation}
Thus, quantum numbers $\kappa^\pm_{j\nu}$ are determined by poles of $G^{j\pm}_{n\alpha,n'\beta}$ as a function of $\kappa_j^\pm$, i.e., by the roots of equation $\sin \kappa_j^\pm N-g_{l,r}^{j}\sin \kappa^\pm_j(N-1)=0$, or \cite{Lyuba1}
\begin{equation}\label{19}
\sin \kappa^\pm_jN=\mp2\cos(\xi_j/2)\sin \kappa^\pm_j(N+1/2),
\end{equation}
which together with Eq.~(\ref{18}) gives energies of 4$N\cal N$ $j,\nu$ levels. Additionally, the graphene spectrum contains two $N$-fold degenerate levels with $E=\pm1$.

In most cases, it is an interval near the Fermi energy, where interesting physics occurs. Whithin this interval and for $N,{\cal N}>>1$, we have \cite{Lyuba3}
\begin{equation} \label{20}
E_c^{(j^*\pm\mu)-}= \left \{
\begin{array}{l}
\sqrt{\mu^2 \Delta^{2}_{_{\rm G}} + \kappa^2/4 },\\
\sqrt{(\mu\mp \frac{1}{3})^2 \Delta^{2}_{_{\rm G}} +\kappa^2/4},  
\end{array} \right.
\end{equation}
where $\Delta_{_{\rm G}}$ = $\sqrt{3}\pi/[2({\cal N}$+1)], $\mu<<\cal N$, and the upper and lower lines refer to an integer and rational value of $2({\cal N}{\rm +}1)/3$, respectively. In the case of integer values of $2({\cal N}{\rm +}1)/3$=$j^*$, say, for $\cal N$=${\cal N}^*$, infinitely long armchair graphene ribbons (GRs) have a metallic spectrum. In contrast, for ${\cal N}$=${\cal N}^*$$\pm1$, the lowest energy mode is $j^*$=$2({\cal N}^*$  $\pm1)/3$, and the spectrum has a band gap ($E_g=2\Delta_{_{\rm G}}/3$), as defined by the lower line in Eq.~(\ref{20}). 

The Green's functions of GRs with the length $N$ and $\cal N$ = ${\cal N}^*$, ${\cal N}^*$+1 and ${\cal N}^*{\rm -1}$ have a pronouncedly different appearance; see Fig.~3. In what follows, we concentrate on energy intervals below (above) the bottom (top) of the second lowest band of metallic GRs, $|E|$$\le$$ \Delta_{_{\rm G}}$, and whithin the band gap of semiconducting GRs, $|E|$$\le$$E_g/2$. For these energies, the Green's functions shown in the mid and right panel in Fig.~3 are accurately reproduced by a single member of expansion (\ref{16}), namely, by the $j^*$th term. To reproduce the energy dependence of the Green's function illustrated in the left panel, two terms are needed, partial Green's functions $j^*$ and $j^*$+1.

\begin{figure}[t]%
\centering
\includegraphics*[width=0.8\linewidth,height=0.62\linewidth]{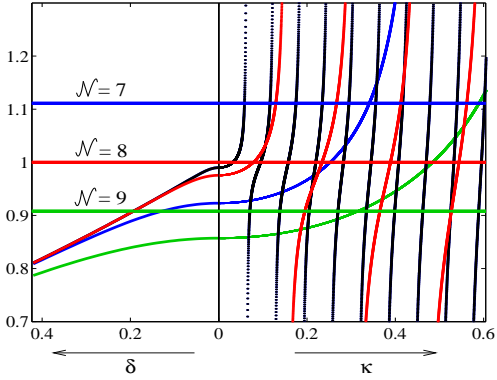}
\caption{Solutions to Eq.~(\ref{19}), shown by intersections of black, red, and blue curves, corresponding to $N$ =50, 20, and 6, respectively, with three horizontal lines $2\cos \xi_{j^*}/2$ for ${\cal N}$ = ${\cal N}^*$ (red),   ${\cal N}^*$-1 (blue), and ${\cal N}^*{\rm +1}$ (green); ${\cal N}^*$=8. Green curve, representing $N$=3 ($<$$\cal N$/2) does not have intersections in the region of imaginary $\kappa=i\delta$; see text. }
\label{Fig4}
\end{figure}

As immediately follows from Eq.~(\ref{20}), the $E$--$\kappa$ relation can be satisfied only by real and only by imaginary ($\kappa$=$i\delta$) values of the wave vector for metallic and semiconducting GRs, respectively. As a result, an exponential factor 
\begin{equation}\label{21}
G^{j^*}_{1 l,Nr}
\sim \exp\left ( -2N\sqrt{(E_g/2)^2-E^2}\right ),
\end{equation}
appears in the Green's function of semiconducting GRs. This factor governs the probability of single-mode electron/hole transmission through a potential barrier; see Ref. \cite{Yu} of this issue. For metallic GRs, the number of poles depends on $N$ as, approximately, $[\sqrt{3}N/({\cal N}{\rm+1)+}1/2]$. For semiconducting GRs with $\cal N$=${\cal N}^*{\rm -1}$, there are no poles at all, but in the case of $\cal N$=${\cal N}^*{\rm +1}$, there is a single pole under the condition $N$$>$$0.9\sqrt{3}$$({\cal N}{\rm +1})\pi{\rm -}1/2$ or, approximately $N$$>$${\cal N}/2$. These peculiarities are illustrated in Fig.~3.

To conclude this report, the spectra and Green's functions of graphene and its building blocks, acenes have been discussed in parallel that provides a deeper insight into the origin and particularities of graphene electronic structure. An exact analytical expression of graphene Green's function is found in terms of an expansion over partial Green's functions. The character of the Green function singularities near the Fermi energy has been examined and shown to be qualitatively different for metallic and semiconducting graphene ribbons.

This work was partly supported by Visby program of the Swedish Institute (SI).

\end{document}